Math. Inequalities and Appl., 1, N4, (1998), 559-563.
\magnification=\magstep1
\hsize=6.5 true in

\baselineskip = 12 pt
\vglue 3 cm
\def\square{\vrule height 8pt width 4pt}
\centerline{INEQUALITIES FOR THE MINIMAL EIGENVALUE  \footnote 
{${}^{}$}{1991 \it Mathematics
Subject Classification. \rm Primary 35J05, 35P15.}
\footnote {${}$}{\it Key words and phrases. \rm Inequalities, estimation 
of eigenvalues, 
perturbation theory.}}

\centerline{OF THE LAPLACIAN IN AN ANNULUS}
\vskip\baselineskip
\centerline{A. G. Ramm}
\centerline{Department of Mathematics, Kansas State University, 
Manhattan, KS 66506-2602, USA}
\centerline{ramm@math.ksu.edu}
\vskip\baselineskip
\centerline{P. N. Shivakumar}
\centerline{Department of Applied Mathematics and Institute of Industrial 
Mathematical Sciences}
\centerline{University of Manitoba, Winnipeg, Manitoba R3T 2N2, Canada}
\centerline{shivaku@cc.umanitoba.ca}
\vskip2\baselineskip

ABSTRACT.  We discuss the behavior of the minimal eigenvalue $\lambda$ 
of the Dirichlet Laplacian in the domain $D_1 \backslash D_2:= D$ 
(an annulus) where $D_1$ is a circular disc and
$D_2 \subset D_1$ is a smaller circular disc.  It is conjectured that the 
minimal eigenvalue $\lambda$ has a maximum value when $D_2$ is a concentric 
disc.  If $h$ is a displacement of the center of the disc 
$D_2$ and $\lambda(h)$ is the corresponding minimal eigenvalue, then 
${d\lambda (h) \over dh} < 0$ so that $\lambda (h)$ is minimal 
when $\partial D_2$ touches $\partial D_1$, where $\partial D$ is the 
boundary of $D$. Numerical results are given to back the conjecture.
Upper and lower bounds are given for $\lambda(h)$.
The above conjecture is proved.
\vskip2\baselineskip
\parskip=12pt

\noindent \bf {1. Introduction}

\parindent=20pt
\rm Let $D_1$ be a disc on $\bf{R} \rm^2$, centered at the origin, of 
radius $1$, $D_2 \subset D_1$ be a disc of radius $a<1$, the center 
$(h,0)$ of which is at the distance $h$ from the origin.
Denote by $\lambda(h)$ the minimal Dirichlet eigenvalue of the Laplacian 
in the annulus $D:=D_h:=D_1 \backslash D_2$.

In this paper the following conjecture is formulated and proved:

\noindent {Conjecture C.} \it The minimal eigenvalue $\lambda(h)$ is a 
monotonically decreasing function of $h$ on the interval 
$0 \leq h \leq 1 -a$.  In particular 
$$\lambda(0) > \lambda(h), \qquad h>0.\leqno(1.1)$$

Let $\dot\lambda:={d\lambda \over dh}$ and let $S$ denote $\partial D_2$, 
the boundary of $D_2$.

The following results are given to back this conjecture:

\noindent \bf Lemma 1. \it One has
$$\dot \lambda = \int_S u^2_N N_1 ds, \leqno (1.2)$$
where $N$ is the unit normal to $S=S_h$ pointing into the annulus $D_h$, 
$N_1$ is the projection of $N$ onto $x_1$-axis, $u_N$ is the normal 
derivative of $u$, and $u(x) = u(x_1,x_2)$ is the normalized in $L^2(D)$ 
eigenfunction corresponding to the first eigenvalue $\lambda$:
$$\Delta u + \lambda u = 0 \qquad in \quad D,\ u= 0 \qquad on \qquad 
\partial D_1 \cup \partial D_2: = \partial D, \leqno (1.3)$$
$$\|u\|_{L^2(D)} = 1. \leqno (1.4)$$

\rm It is argued at the end of Section 2 that
$$\dot \lambda < 0 \qquad if \qquad 0 < h < 1-a. \leqno (1.5)$$

In Lemma 2 below we give upper and lower bounds (1.6) for $\lambda(h)$.  
These bounds are practically convenient, especially for small $h$.

Let $D(r)$ be the disc $|x|\leq r, \quad \mu(r)$ be the first Dirichlet 
eigenvalue of the Laplacian in $D_1\setminus D_1(r)$.
In Section 3 inequality (1.5) is
illustrated by the
numerical results in $D_1 \backslash D(r)$.

\noindent \bf {Lemma 2.} \it One has
$$\mu(a-h) < \lambda(h) <\mu(a+h), \qquad 0 < h < 1 - a,\quad h<a. \leqno
(1.6)$$
\rm In section 2 proofs are given and the conjecture is proved.

\noindent \bf {2. Proofs}.

\noindent \it Proof of Lemma 2. \rm Lemma 2 is an immediate consequence 
of the variational principle for $\lambda$ since 
$D_1\setminus D(a+h) \subset D_h \subset D_1\setminus D(a-h)$. 
Note that $\mu(b),$ $a \leq b < 1$, can be calculated efficiently.  
Indeed, by symmetry the first eigenfunction $\phi$ of the Dirichlet 
Laplacian in $D_1 \backslash D(b)$ depends on the radial variable $r=|x|$ 
only, and solves the problem
$$\phi'' + {1 \over r} \phi ' + \mu \phi = 0, \qquad b \leq r \leq 1; \qquad
\phi(b) = \phi(1) = 0. \leqno (2.1)$$
Thus
$$\phi=c_1J_0 (\sqrt{\mu } r)+c_2N_0(\sqrt{\mu } r), \leqno (2.2)$$
where $J_0$ and $N_0$ are the Bessel functions, and $c_1$, $c_2$ are 
constants.  The boundary conditions (2.1) are satisfied if $\mu=\mu(b) 
>0$ is a positive root of the equation:
$$J_0(\sqrt{\mu }b) N_0 (\sqrt{\mu})-J_0 (\sqrt {\mu})N_0 (\sqrt {\mu }
b)=0. \leqno (2.3)$$
The smallest positive root $\mu=\mu(b)$ of (2.3) is the desired first 
eigenvalue of the Dirichlet Laplacian in $D_1 \backslash D(b)$. 
Equation (2.3) can 
be solved numerically.  This makes (1.6) an efficient estimate of 
$\lambda(h)$, especially for small $h>0$.

\noindent \square

\noindent \it Proof of Lemma 1. \rm We use the known technique based on 
the domain derivative [1].

It is known that $\lambda(h)$ is continuously differentiable with respect 
to $h\ [2]$.  Let $\dot u = {du\over dh}$, where $u$ solves (1.3)-(1.4).  
Differentiate the equation and the boundary condition (1.3) with 
respect to $h$ and get
$$\Delta \dot u + \lambda \dot u = - \dot \lambda u \qquad in \qquad D = 
D_h, \leqno (2.4)$$
$$\dot u + u_NN_1 = 0 \  on \  S = S_h. \leqno (2.5)$$
Multiply (2.4) by $u$, (1.3) by $\dot u$, subtract, integrate over 
$D=D_h$, use Green's formula, and (2.5) and get:
$$\dot\lambda \int_D u^2dx = \int_S (u \dot u_N - \dot u u_N)ds = \int_S 
u_N^2 N_1ds.\leqno (2.6)$$
From (2.6) and (1.4) one gets (1.2).  Lemma 1 is proved.

\noindent \square

It follows from (1.2) by symmetry that $\dot \lambda (0) = 0$.  Indeed, 
if $h=0$, then $u_N^2|_{S_0}$ = const by symmetry, and $\int_{S_0} N_1ds 
= 0$.

If $h>0$, then $u_N^2$ on the half circle $S_h^+$, the part of the 
boundary of $S_h$ which is closer to $\partial D_1$, is likely to be less 
than on the other half $S_h^-$ of $S_h$, while $N_1>0$ on $S_h^+$ and 
$N_1 < 0$ on $S_h^-$. Moreover, $|N_1|$ is the same at the symmetric 
points of $S_h^+$ and $S_h^-$, where the axis of symmetry is the vertical 
diameter of $D_2$.  Therefore one expects 
$\dot \lambda (h) <0$ for $h>0$, which is the conjecture $C$.

Let us prove that the above argument is indeed valid.
What we wish to prove is the inequality for the normal derivative
$u_N$ mentioned above. 

The following argument completes the proof of the Conjecture (C).
This argument was communicated to AGR by Professor M.Ashbaugh.
 Consider the reflection of the part of the domain
which is situated to the right of the 
vertical line passing through the center of the smaller disc
with respect to this line.
Let $D_h$ denote the domain symmetric with respect to this line $\ell$
and $v$ denote the function equal to $u$ to the right of $\ell$,
and equal to $w$ to the left of $\ell$. Here $w(x,y)=u(x,-y)$,
where the $y$-axis is the line $\ell$.  
By the maximum principle one has $u>v$ on the part of the boundary
of $D_h$ which lies to the left of $\ell$ and, by the Hopf lemma
(strong maximum principle), it follows that $u_N>v_N$ on this part 
of the boundary of $D_h$. This is the desired inequality since
$v=u$ to the right of  $\ell$.

\noindent \bf{3. Numerical Results} 

\rm We use a finite element method to calculate $u_N^2$ at a number of 
nodal points $\phi$ on $\partial D_2$, where $\phi$ is the angle 
between the radial line at the positive 
x-axis.  Due to symmetry, it is sufficient to consider $0 \leq \phi \leq 
\pi$.  The following tables give values for $u_N^2$ for various values of 
$h$ and $\phi$.  The last row gives $\lambda(h)$ for different values of $h$. 

\vskip\baselineskip
\centerline {Table 1}
\centerline {Values for $u_N^2$}
\vskip\baselineskip

\centerline {$a=0.1 \qquad \qquad \lambda (0) = 10.98324859$}
\vskip\baselineskip
\settabs 5 \columns

\+ & $h=0.1$ & $h=0.3$ & $h=0.6$ & $h=0.8$ \cr
\+ \cr
\+ $\phi$ \cr
\+ \cr

\+ $0^\circ$ & $0.18340156$ & $0.08997194$ & $0.03502936$ & $0.00538128$ \cr
\+ $15^\circ$ & $0.18586555$ & $0.09354750$ & $0.03875921$ & $0.00792279$ \cr
\+ $30^\circ$ & $0.19312533$ & $0.10408993$ & $0.04977909$ & $0.01615017$ \cr
\+ $45^\circ$ & $0.20478642$ & $0.12105508$ & $0.06745736$ & $0.03118122$ \cr
\+ $60^\circ$ & $0.22019869$ & $0.14357017$ & $0.09052611$ & $0.05294918$ \cr
\+ $75^\circ$ & $0.23846941$ & $0.17048691$ & $0.11728631$ & $0.07901455$ \cr
\+ $90^\circ$ & $0.25848609$ & $0.20042583$ & $0.14624640$ & $0.10706183$ \cr
\+ $105^\circ$ & $0.27895498$ & $0.23176494$ & $0.17645804$ & 
$0.13678539$ \cr
\+ $120^\circ$ & $0.29846292$ & $0.26256868$ & $0.20707716$ 
&Ê$0.16793719$ \cr
\+ $135^\circ$ & $0.31556947$ & $0.29053976$ & $0.23653390$ & 
$0.19964213$ \cr
\+ $150^\circ$ & $0.32892971$ & $0.31313057$ & $0.26197403$ & 
$0.22879649$ \cr
\+ $165^\circ$ & $0.33743644$ & $0.32789921$ & $0.27954557$ 
&Ê$0.24990529$ \cr
\+ $180^\circ$ & $0.34035750$ & $0.33304454$ & $0.28585725$ & 
$0.25766770$Ê\cr 
\vskip\baselineskip
\+ $\lambda(h)$ & $10.51624800$ & $8.76956649$ & $6.91928150$ & 
$6.21431318$ \cr

\vskip.5in
\centerline{Table 2}
\centerline{Values for $u_N^2$}
\vskip\baselineskip
\centerline{$a=0.3 \qquad \qquad \lambda(0) = 19.46950428$}
\vskip\baselineskip
\settabs 4 \columns

\+ & $h=0.1$ & $h=0.3$ & $h=0.6$ \cr
\+ \cr
\+ $\phi$ \cr
\+ \cr
\+ $0^\circ$ & $0.04651448$ & $0.00601084$ & $0.00006665$ \cr
\+ $15^\circ$ & $0.05078040$ & $0.00792264$ & $0.00029224$ \cr 
\+ $30^\circ$ & $0.06389146$ & $0.01432651$ & $0.00162487$ \cr
\+ $45^\circ$ & $0.08665951$ & $0.02711431$ & $0.00616138$ \cr
\+ $60^\circ$ & $0.12001996$ & $0.04901522$ & $0.01734345$ \cr
\+ $75^\circ$ & $0.16444947$ & $0.08285892$ & $0.03916871$ \cr
\+ $90^\circ$ & $0.21927390$ & $0.13049149$ & $0.07481155$ \cr
\+ $105^\circ$ & $0.28204163$ & $0.19150347$ & $0.12521694$ \cr
\+ $120^\circ$ & $0.34820007$ & $0.26211532$ & $0.18784387$ \cr
\+ $135^\circ$ & $0.41130766$ & $0.33475001$ & $0.25580537$ \cr
\+ $150^\circ$ & $0.46389778$ & $0.39885669$ & $0.31827254$ \cr
\+ $165^\circ$ & $0.49888764$ & $0.44319924$ & $0.36272535$ \cr
\+ $180^\circ$ & $0.51117180$ & $0.45907590$ & $0.37887932$ \cr 
\vskip\baselineskip
\+ $\lambda(h)$ & $17.00607073$ & $12.31240018$ & $8.54494014$ \cr

\vfill\eject
\centerline{Table 3}
\centerline{Values for $u_N^2$}
\vskip\baselineskip
\centerline{$a=0.6 \qquad \qquad \lambda(0) = 61.2854372$}
\vskip\baselineskip
\settabs 3 \columns

\+  & $h=0.1$ & $h=0.3$ \cr
\+ \cr
\+ $\phi$ \cr
\+ \cr
\+ $0^\circ$ & $0.00010994$ & $0.00000018$ \cr
\+ $15^\circ$ & $0.00025775$ & $0.00000144$ \cr
\+ $30^\circ$ & $0.00101252$ & $0.00002268$ \cr
\+ $45^\circ$ & $0.00370221$ & $0.00026580$ \cr
\+ $60^\circ$ & $0.01190759$ & $0.00195778$ \cr
\+ $75^\circ$ & $0.03332159$ & $0.00947178$ \cr
\+ $90^\circ$ & $0.08086609$ & $0.03287792$ \cr
\+ $105^\circ$ & $0.17026477$ & $0.08782665$ \cr
\+ $120^\circ$ & $0.32267905$ & $0.18896048$ \cr
\+ $135^\circ$ & $0.49728793$ & $0.33653240$ \cr
\+ $150^\circ$ & $0.69311417$ & $0.50402714$ \cr
\+ $165^\circ$ & $0.84533543$ & $0.64040281$ \cr
\+ $180^\circ$ & $0.90307061$ & $0.69330938$ \cr 
\vskip\baselineskip
\+ $\lambda(h)$ & $42.71463081$ & $23.79696055$ \cr

\noindent  In all the cases above, $u_N^2$ increases in value as $\phi$ 
increases from zero to $\pi$, thereby confirming that 
$\dot\lambda < 0$ (see formula (2.6)).  
From the above tables we also note that for fixed 
$a$, $\lambda(h)$ is a decreasing function of $h$,
and that $\lambda(h) < \lambda(0)$ for $h > 0$ thus confirming the 
Conjecture C.

\vskip\baselineskip

\centerline{REFERENCES}

\item {1.} J. Sokolowski, J. Zolezio, \it Introduction to shape 
optimization, \rm Springer Verlag, Berlin, 1992

\item {2.} T. Kato, \it Perturbation theory for linear operators, \rm 
Springer Verlag, Berlin, 1966.

\bye